\pdfoutput=1
\documentclass[letterpaper]{JINST}
\usepackage{url}

\newcommand{\gm}{$g$\,$-$\,$2$}
\newcommand{\pb}{PbF$_2$}

\title{Design and performance of SiPM-based readout of \pb~crystals
for high-rate, precision timing applications}
\author{%
  J.~Kaspar$^a$\thanks{Corresponding author.},
  A.\,T.~Fienberg$^a$,
  D.\,W.~Hertzog$^a$,
  M.\,A.~Huehn$^a$,
  P.\,~Kammel$^a$,
  K.\,S.~Khaw$^a$,
  D.\,A.~Peterson$^a$,
  M.\,W.~Smith$^a$,
  T.\,D.~Van\,Wechel$^a$,
  A.\,Chapelain$^b$,
  L.\,K.\,Gibbons$^b$,
  D.\,A.\,Sweigart$^b$,
  C.\,Ferrari$^{c,d}$,
  A.\,Fioretti$^{c,d}$,
  C.\,Gabbanini$^{c,d}$,
  G.\,Venanzoni$^c$,
  M.\,Iacovacci$^{e,f}$,
  S.\,Mastroianni$^e$,
  K.\,Giovanetti$^g$,
  W.\,Gohn$^f$,
  T.\,Gorringe$^h$,
  D.\,Pocanic$^i$
  \\
\llap{$^a$}University of Washington, Department of Physics, Box 351560, Seattle, WA 98195, USA\\
\llap{$^b$}Cornell University, Department of Physics, 511 Clark Hall, Ithaca, NY 14853, USA\\
\llap{$^c$}Laboratori Nazionali Frascati dell' INFN, Via E. Fermi 40, 00044 Frascati, Italy\\
\llap{$^d$}Istituto Nazionale di Ottica del C.N.R., UOS Pisa, Via Moruzzi 1, 56124 Pisa, Italy\\
\llap{$^e$}INFN, Sezione di Napoli, Ed. 6 Via Cintia, 80126 Napoli, Italy\\
\llap{$^f$}Universit\`a di Napoli, Corso Umberto I 40, 80138 Napoli, Italy\\
\llap{$^g$}James Madison University, Department of Physics and Astronomy,  901 Carrier Dr, Harrisonburg, VA 22807, USA\\
\llap{$^h$}University of Kentucky, Department of Physics and Astronomy, 505 Rose Street, Lexington, KY 40506, USA\\
\llap{$^i$}University of Virginia, Department of Physics, 382 McCormick Rd, Charlottesville, VA 22904, USA\\

  E-mail: \email{kaspar@uw.edu}}

\abstract{%
  We have developed a custom amplifier board coupled to a large-format 16-channel Hamamatsu silicon photomulplier device for use as the light sensor for the electromagnetic calorimeters in the Muon \gm~ experiment at Fermilab.  The calorimeter absorber is an array of lead-fluoride crystals, which produces short-duration Cherenkov light.  The detector sits in the high magnetic field of the muon storage ring.  The SiPMs selected, and their accompanying custom electronics, must preserve the short pulse shape, have high quantum efficiency, be non-magnetic, exhibit gain stability under varying rate conditions, and cover a fairly large fraction of the crystal exit surface area.  We describe an optimized design that employs the new-generation of thru-silicon via devices. The performance is documented in a series of bench and beam tests.
}


\begin{document}

\section{Introduction}

The rapid development of, and subsequent improvements in, commercial pixelated silicon photo-diode sensors operated in Geiger mode --- silicon photomultipliers, or SiPMs\footnote{Many names exist, but the community is largely converging on SiPM as a useful pneumonic.  The Hamamatsu products described herein are called Multi-Pixel Photon Counters, or MPPCs, by the company.} --- has led to a vast array of end-use applications in the physical sciences and in medical imaging~\cite{Renker:2006ay}.
High-energy and nuclear physicists increasingly are turning to the use of SiPMs in applications where small photomultiplier tubes (PMTs) would have otherwise been used (e.g., Refs.~\cite{Smith:2016yvt,Ieki:2016mox}). This is particularly well motivated when space for the light sensor is constrained --- SiPMs and their readout boards can be quite compact --- and also when the detector must reside in a high magnetic field environment, as SiPMs performance is unaffected by multi-Tesla fields.

SiPMs have high quantum efficiencies across a broad range of wavelengths, which are typically emitted by scintillation detectors or even Cherenkov light sources.  Because of the avalanche properties' nature of single pixels within a device, the intrinsic rise time is fast, making them ideal for time-of-flight counters and other applications having a premium on timing performance.  In many applications originally using SiPMs, the emphasis has been on few photo-electron (PE) counting situations, where the single PE sensitivity is critical and easily exceeds that of PMTs. Impressive few PE resolution is typically a feature of commercial catalogs documenting performance.  In contrast, here we describe the use of SiPMs in the many hundreds to thousands of PE regime associated with large light output situations such as calorimetry.  As described below, even here, the SiPM holds many advantages over a PMT.

The dream that a SiPM can simply replace a PMT at a fraction of the cost, on the other hand, is naive. While costs have indeed fallen considerably, great care must be taken to enlarge the photo-sensitive surface and maintain the intrinsic performance of the smaller, core devices, which are usually only a few square mm in size. At present, the user can only purchase pre-packaged arrays\footnote{e.g., $4\times4$, $8\times8$, and $16\times16$ arrays of 9\,mm$^2$ units are now available from Hamamatsu.} from the manufacturer as single devices or build custom arrays from chips that can be mounted edge-to-edge with very little dead space between them.  In either case, the sub-units, or channels, must be gain matched and summed to produce the output of an effective single element.

The gain of a SiPM depends sensitively on the overvoltage above the Geiger-mode breakdown threshold, $V_{bd}$.  Since $V_{bd}$ depends on temperature, gain is intrinsically linked to bias voltage supply stability {\em and} to external  temperature stability.  The typical dependence of $V_{bd}$ on temperature for recent Hamamatsu devices is 60\,mV per $^\circ$C at the recommended operational overvoltage.  The gain sensitivity at the overvoltage operational point is $\sim0.2\%$ per mV.  Because of this, the mechanical housing, with some sort of cooling circuit or stabilization, and the bias control system, possibly with feedback based on online or offline SiPM
temperature readout, must be carefully considered.  For large-area, multi-channel devices, attention to the summing and pulse-shape electronics is critical and non-trivial. Finally, if the application involves high rate and potential pulse pile-up, special attention must also be paid to the bias voltage supply and its interface to the SiPM in order to provide an adequate current source to rapidly replenish the extracted charge from fired pixels.

Despite existing equivalent circuit models to guide the design, fine-tuning of the amplifier circuit owing to finite tiny capacitances, inductances, and resistance in circuit elements and board traces can be critical to achieve an optimized pulse shape.   Our application goal for a light sensor requires a {\em PMT-like} performance with respect to pulse shape and two-pulse resolution as well as a {\em better-than-PMT} stability against high rate.  In the report that follows, we demonstrate a design that meets these goals, which we believe will be of general interest to the community.

\section{Application-Specific Design Constraints}

For context, we  briefly describe the application-specific requirements of the Fermilab Muon \gm\ experiment. Sixteen bunches of 3.1~GeV/$c$ polarized positive muons will be injected into the muon storage ring~\cite{Danby:2001eh,Grange:2015fou} every 1.4\,s.  As the muons decay ($\gamma\tau_\mu = 64.4~\mu$s), the emitted positrons, having any energy up to $\sim3.1$~GeV, will curl towards the ring interior where they might strike one of the 24 electromagnetic calorimeter stations that are located at regular intervals immediately adjacent to the storage ring volume.  The magnetic field at the larger radius of the calorimeter is 1.45\,T, and it drops to 0.8\,T over the 225~mm width of the detector.  The calorimeter is segmented into six rows and nine columns of $25 \times 25 \times 140$~mm$^3$ lead-fluoride (PbF$_2$) crystals. Each crystal is read out by a single monolithic SiPM.

The calorimeter is required to provide information
for determining the energy and time-of-arrival of each decay positron with the following emphases:
\begin{itemize}
\item The time resolution, extracted from a fit of the SiPM pulse, must be better than 100\,ps for positrons having energy greater than 100\,MeV.
\item The calorimeter must be able to reliably resolve two showers
    by temporal separation with time separations greater than 5\,ns, and it must accurately assign correct energies to the two pulses.
\item The energy resolution of the reconstructed positron energy
    summed across calorimeter segments must be better than 5\,\% at 2\,GeV.
\item The maximally allowed, uncorrected, gain change is
    $\delta G/G < 0.04$\,\% over the 700\,$\mu$s time period while the rate reduces by four orders of magnitude, starting with an initial rate that can exceed 1\,MHz.
\item The analog and digital signal chains must preserve the pulse shape of the Cherenkov light collection.
\item The calorimeter, readout, and cabling all must reside adjacent to a highly uniform magnetic field, and the components must not perturb this field.
\end{itemize}

In Ref.~\cite{Fienberg:2014kka}, we documented many calorimeter-specific performance tests that have led to the final calorimeter design. That test employed an earlier version of the SiPM circuit design and used simplified digitization electronics to record the data. Here we describe the final SiPM design and performance and selected beam-test results using the custom 800~MSPS and 12-bit depth digitizers~\cite{Chapelain:2015esj,Grange:2015fou}, which will be used in the Muon \gm~ experiment.

\section{SiPM Selection}
Because multi-particle pile-up separation is a dominant systematic effect in our high-rate application, we looked for devices delivering a narrow pulse shape without additional shaping or clipping. The type of silicon substrate and carrier mobility contribute to the pulse shape. Devices with better carrier mobility deliver narrower pulses but typically require a higher bias voltage~\cite{Dinu2009423}.
Because the number of Cherenkov photons in our application is greater than that needed to fulfill our requirements on energy resolution, the absolute quantum efficiency of the SiPM was not among the critical parameters.

While the market has evolved considerably since we began our development,
the Hamamatsu surface-mount 16-channel SiPM used here
is a fairly typical modern device appropriate for our development. The 16 channels are packaged as a single $4\times4$ array of 3\,$\times$\,3\,mm$^2$ discrete channels with solder pads on the back.\footnote{Model number S12642-4040PA-50.} The net 12\,$\times$\,12\,mm$^2$ active area is similar to that of a 19-mm diameter photomultiplier, such as the Hamamatsu R750. The 50\,$\mu$m pitch selected yields a total of 57,344 pixels.\footnote{A $4\times4$ pixel sized area centered on each channel is used for the through-silicon via.}

In this newer generation of Hamamatsu SiPMs, the original poly-crystalline silicon quenching resistor was replaced with a Ni-based resistor\footnote{The Ni fraction is tiny; we evaluated the magnetic perturbation of this device in our test magnet, and the effect of Ni creates a negligible perturbation.}, which is less sensitive to temperature variance.   The connection to the summing board uses through-silicon vias to avoid parasitic inductances of the finite length traces, which are in series with the output current pulse.
Optical trenches between pixels minimize cross-talk, and, finally, the use of high-purity silicon wafers minimizes dark current and after-pulsing and increases the charge delivered by a fired pixel.

A photon striking a pixel can cause an avalanche, initiating current flow.  This current is merged with that from the other struck pixels in the same channel. The 16 channels are independent, except that they share the same applied bias voltage. To create a large single active channel --- a true PMT-like device --- requires a custom summing board to add the individual contributions from the channels, preserving the attractive signal characteristics of the smaller units.

After an avalanche is initiated in a pixel, the quenching resistor arrests the flow of current by dropping the effective bias below the Geiger threshold.  The pixel recovers with a time constant typically on the order of 10\,ns. Those pixels that had not been struck remain ready for an event.

In general, good linearity is achieved when the number of pixels in the device, $N_{device}$, greatly exceeds the highest expected number that might fire for a given event. When two photons strike the same pixel, only a single avalanche can result; the response is intrinsically non-linear.
In our application, a maximum of several thousand pixels will fire for the highest-energy positron shower.  The $\sim 5\,\%$ pixel occupancy fraction is in the near-linear regime.  For higher photon count applications, one can reliably estimate the number of pixels, $N_{pixels}$, that would have fired if the device had an infinite number of pixels available,
\begin{equation}
N_{pixels} = -N_{device} \ln\left(1 - \frac{N_{fired}}{N_{device}}\right). \label{eq:fired}
\end{equation}
This conversion requires only that the user has calibrated the output pulse integral/height to determine the fired pixel count, $N_{fired}$, a straight-forward procedure described in Section~\ref{sc:sc-calib}.

Hamamatsu SiPMs require a stable bias voltage in the range of 65 -- 70\,V.  For our application, involving nearly 1300 elements, we grouped 12 -- 16 SiPMs having common manufacturer specified breakdown voltages, to a single bias supply. A stable and nearly identical operating temperature is maintained on these grouped SiPMs by flowing forced outside air across the amplifier board's heat sinks, which are enclosed in a common light-tight detector housing. Accommodation for temperature variances of the group can be made by small bias voltage adjustments if necessary.

\section{Electronics Design}

Several approaches can be used to read out the charge
from 16 channels and sum them into a final pulse output. One can
use a common shunt resistor to create a voltage amplifier, or construct
a transimpedance (current) amplifier based on either discrete or integrated electronics.

The advantages of the voltage amplifier design are its simplicity,
robustness, and ease of simulation, which make it particularly
suitable for the characterization of an unknown SiPM.
However the shunt resistor in the circuit is much larger than the transimpedance design.  Therefore, the pulse width due to the RC time constant is wider.  Furthermore, the bias voltage depends on the current drawn, making the gain dependent on the signal size.
This may not present a problem for time-of-flight applications or
for counting single photons.

For our application in calorimetry ---  where several thousand
pixels can fire in a single event --- a transimpedance amplifier is a better choice. The
arrival time dispersion of the Cherenkov photons at the SiPM is typically around 1\,ns,
and the variance of this distribution --- i.e., the width of the pulse --- provides
critical information on multi-particle pileup and, to a lesser degree, the
incident angle of the incoming particle. The width variance of the input light is preserved
in the output pulse shape. The amplifier does not shape the pulse beyond the
limits of its bandwidth.

Practical factors also enter into the design considerations.  In our
application, the footprint of the printed circuit board (PCB) on which the SiPM is mounted must be less
than $25 \times 25\,$mm$^2$.  The system as a whole requires remote control of the variable gain amplifier, readback of an on-board probe, and local EEPROM storage of a detector identifier and gain settings.
Owing to space restrictions, a single multi-component HDMI connector is used to convey $\pm$5\,V,
the bias voltage, and the communication protocol.
Finally, to match the input requirement of the custom 800\,MSPS digitizer, a differential signal with a 2.5\,V DC common-mode voltage is needed.

\subsection{Spice Model}

Designing the summing circuit requires a model of the SiPM device
itself.  We started by using a SPICE model\footnote{http://bwrcs.eecs.berkeley.edu/Classes/IcBook/SPICE/} incorporating an electronic
equivalent of the SiPM.  Variants have been discussed in
Refs.~\cite{Corsi:2007zz,Seifert2009}.
Following Seifert, the parameters used to describe
one of the 16 channels having 3584 microcells, each of
50\,$\mu$m pitch, are listed in Table~\ref{tb:spicemodel}.

\begin{table} \center
\caption{Parameters used in the SPICE model to describe one of the 16 channels in the device}\label{tb:spicemodel}
\begin{tabular}{l|l}
  \hline
  Description & Value \\
  \hline\hline
  Capacitance of the reverse-biased diode pixel & $C_d$ = 55.8\,fF, \\
  Parasitic bulk capacitance & $C_g$ = 50\,pF \\
  Total measured capacitance of the channel & 250\,pF \\
  Effective series resistance of the avalanche & $R_d$ = 1\,k$\Omega$ \\
  Resistance of the Ni based quench resistor & $R_q$ = 100\,k$\Omega$ \\
  Parasitic stray capacitance of the quench resistor & $C_q$ = 3.4\,fF\\
  Pixels fired in each of the 16 channels of the array & $N_c$ = 200 \\
  \hline
\end{tabular}
\end{table}

Two principle changes have been made to the current generation of Hamamatsu SiPM products
compared to those modeled originally in the literature. They are
fabricated using higher purity silicon wafers, and the quenching resistor now uses
a Ni-based one rather than poly-crystalline silicon.

The 16-channel SiPM has a common bias line, but the $3 \times
3$\,mm$^2$ individual channels are isolated electrically; they
represent 16 truly independent channels.  The goal is to sum the 16 channels
to create an effectively larger device, while preserving their timing
and pulse-shape characteristics. Three
options considered include: 1) a simple sum of all 16 channels in one
stage,  2) an amplification of each of the 16 channels separately, followed by adding the outputs, and, 3) a multi-staged approach, which we adopted and will describe next.
The selection of 3) reduces the capacitance four-fold compared to 1) and, therefore, reduces the RC time constant controlling the pulse width associated with the pixel recover time.  Option 2 would be better, but it is difficult to implement because of the increased local heat generation and component space restrictions.

The SPICE model and board prototypes were developed in an iterative way with 12 versions of prototype boards in total. The performance of each version was evaluated with a variable width pulse generator driving an LED, and a narrow width laser. The pulse generator, Avtech AVPP-1A-B, provided  a pulse width in the range from 0.5\,ns to 3.0\,ns. The pulsed laser, Picoquant LDH-P-C-405 driven by a PDL 200-B, was used to evaluate pulse-shape fluctuations and high-rate stability. The feedback from the prototype tests was used mainly to tune parasitic impedances of the first-stage components and the PCB material.

\begin{figure}[t]
    \includegraphics[width=0.95\textwidth]{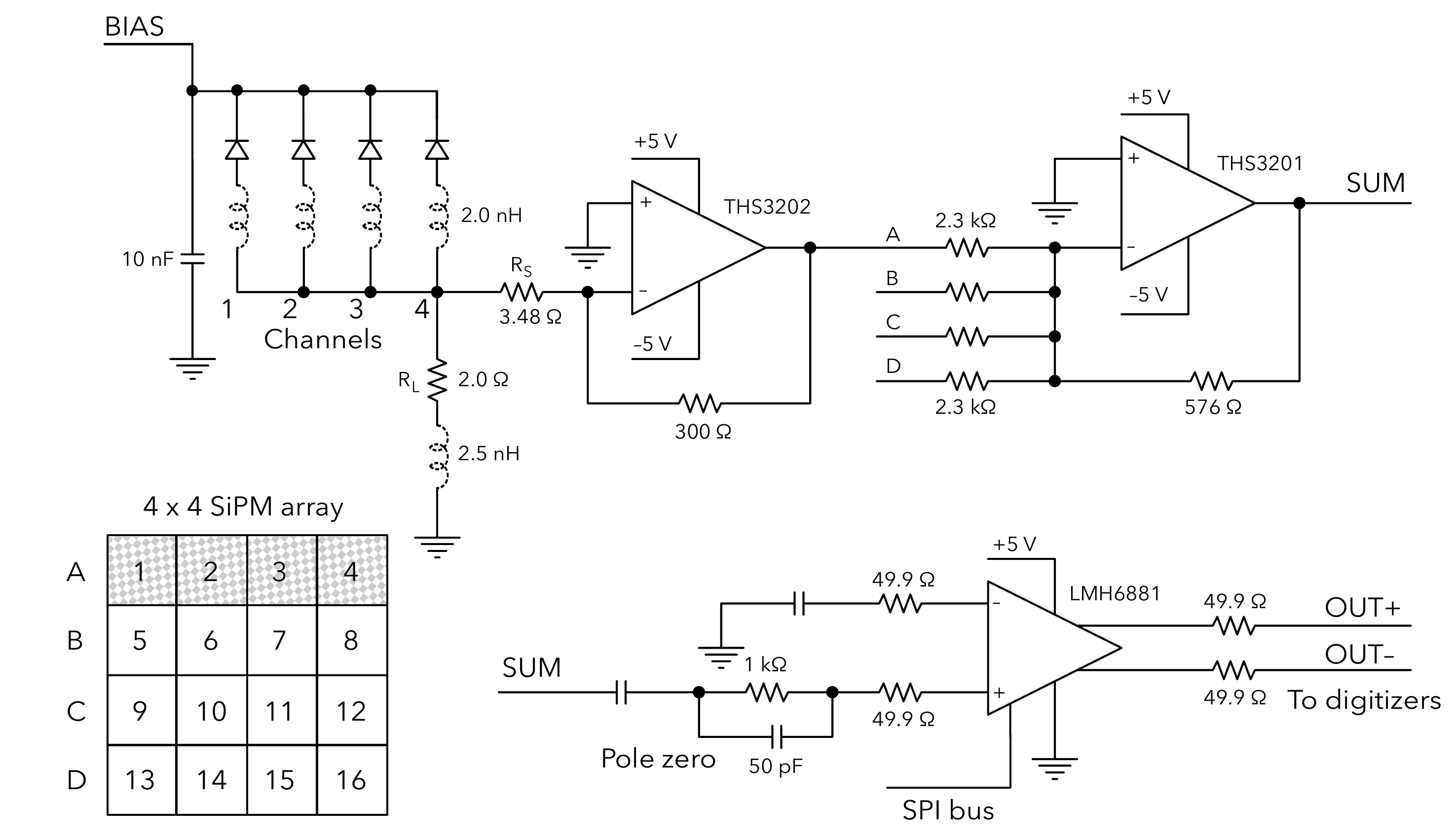}
\caption{Simplified schematic of the amplifier board. The 16 channels of the SiPM share a common bias voltage. The circuit proceeds from the top left by first summing four channels within a row and then summing the rows. A final output stage, which employs a variable gain amplifier, delivers the differential signal to external waveform digitizers. The dashed inductors are present only in the SPICE simulation and represent parasitic properties of the components and PCB. \label{fg:schematic}}
\end{figure}

\subsection{Final Design}
The final schematic, shown in Fig.~\ref{fg:schematic} is based on the concept of a multi-staged amplifier.  At the inverting input of the first stage, current pulses from four SiPM channels are summed together and converted into voltage pulses in a fixed-gain transimpedance amplifier, which is operated at a constant gain of 300.  This first stage is designed around a THS3202 2-GHz bandwidth current feedback op-amp.   The impedance looking into the input of the op-amp is very low --- less than 100\,m$\Omega$ --- at frequencies below 1\,MHz.  Above 1\,MHz, the input impedance has a rising characteristic as a function of increasing frequency.  The net input impedance approximates a 7\,nH inductor in series with 50\,m$\Omega$.    If the SiPM channels were connected directly to the op-amp input, the input inductance would resonate with the SiPM's capacitance $C$, resulting in high-frequency oscillations.  The resistor $R_S = 3.48\,\Omega$, which is placed in series between the common anode side of a group of four SiPMs and the op-amp input, dampens this resonant behavior.  Another resonance exists at a lower frequency owing to the inductance of the printed circuit board traces and interconnections, and the SiPM capacitance.   This resonance does not cause oscillation, but instead adds a damped sine wave to the exponential tail of the SiPM pulse.  This resonance is dampened by resistor $R_L = 2.0\,\Omega$.

In the second stage, the four partial row sums are added together using a THS3201 op-amp operated at a voltage gain of 0.25.

The final stage is designed around a LMH6881 digitally controlled variable gain differential amplifier, which has differential inputs and outputs.   The output of this stage drives a shielded twisted pair connected to a waveform digitizer, which has a 100-$\Omega$ differential input impedance with a common mode voltage of 2.5\,V.  The amplifier's common-mode output voltage is set by a resistive voltage divider of the $+5$\,V power supply connected to a bias set pin that is not shown in the simplified schematic.  The inputs of the differential amplifier are AC-coupled through 0.1\,$\mu$F capacitors in series with 49.9\,$\Omega$ resistors.  The negative input is AC-coupled to ground while the positive input is AC-coupled to the output of the summing amplifier.

A pole-zero network is inserted in series with the positive input, consisting of a 50\,pF capacitor in parallel with a
1\,k$\Omega$ resistor.  This reduces the width of the output pulse by
canceling out the exponential tail of the output pulse, but it preserves
the meaningful sensitivity to the pulse width that corresponds to the arrival time of
the photons impinging on the SiPM active surface. The pole-zero network also removes the unwanted artifacts of AC coupling, serving as a weak baseline restorer; and it
improves the overall stability of the signal chain.  The SiPM chip
inherently behaves as a capacitor itself, and the additional AC
coupling between the amplifiers does not introduce a new problem.  The
solution preserves the pulse shape as accurately as practically
possible.

\begin{figure}[t]
    \includegraphics[width=0.95\textwidth]{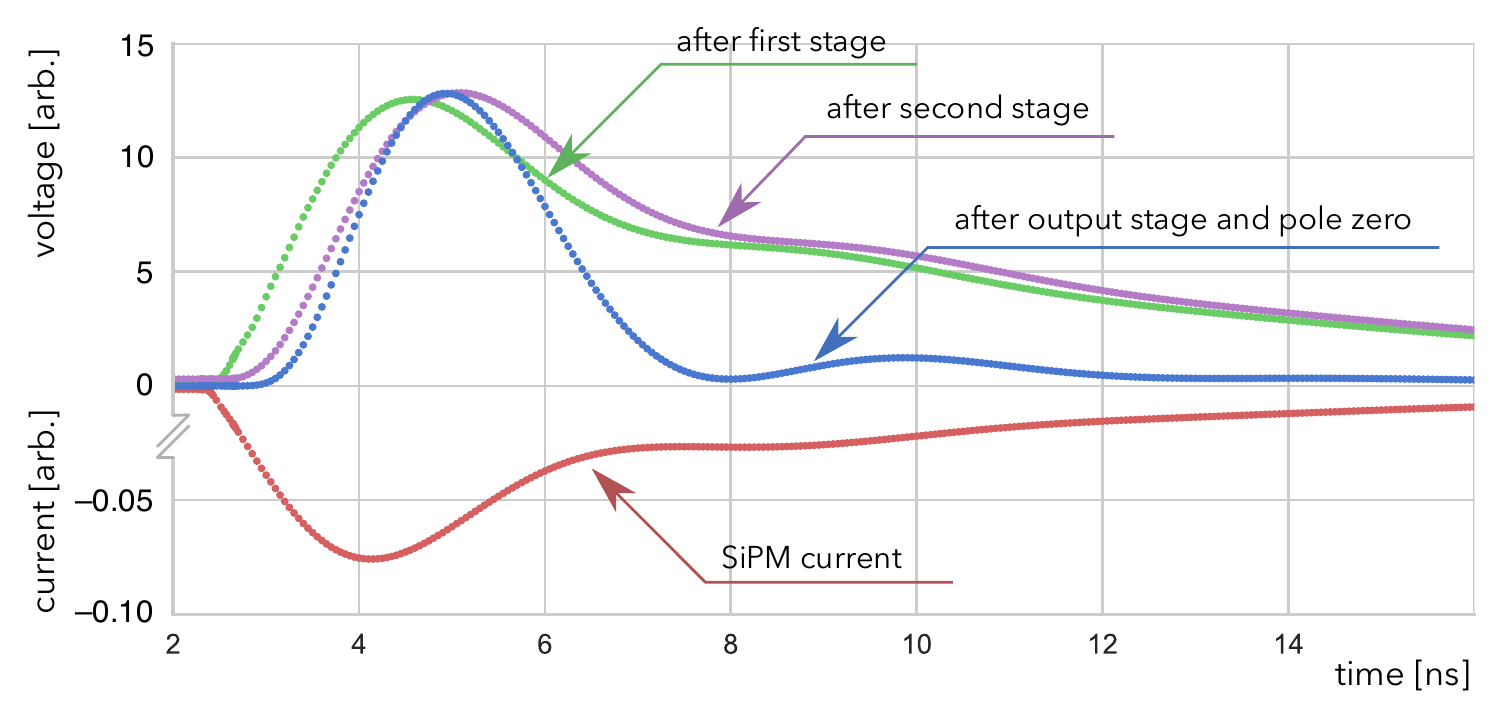}
\caption{Upper three traces: SPICE simulation of the SiPM pulse evolution as the pulse undergoes the three amplification stages on a SiPM board. The amplitudes of the pulses are arbitrarily scaled for an easier comparison of the pulse shapes. Lower trace:  SiPM current in the model.\label{fg:spice-pulses}}
\end{figure}

Figure~\ref{fg:spice-pulses} shows the SPICE model predictions of the absolute value of the pulse shape after the first, second, and final stages of the amplifier board.  We emphasize that these are the shapes the model produces; the actual pulse shape after the second stage is narrower, and the final output shape can be compared directly to what is seen in Fig.~\ref{fg:pulses}. The lower trace in Fig.~\ref{fg:spice-pulses} shows the current in the SiPM in the context of the SPICE model.

The gain of the differential output amplifier is set by a BeagleBone microcomputer over an SPI data bus.  The gain is adjustable over a 20\,dB range in 0.25\,dB steps.
A 1\,Vpp amplitude at the output of the amplifier board corresponds to
approximately 4000 pixels fired across all 16 channels.

\begin{figure}[t]
    \includegraphics[width=0.95\textwidth]{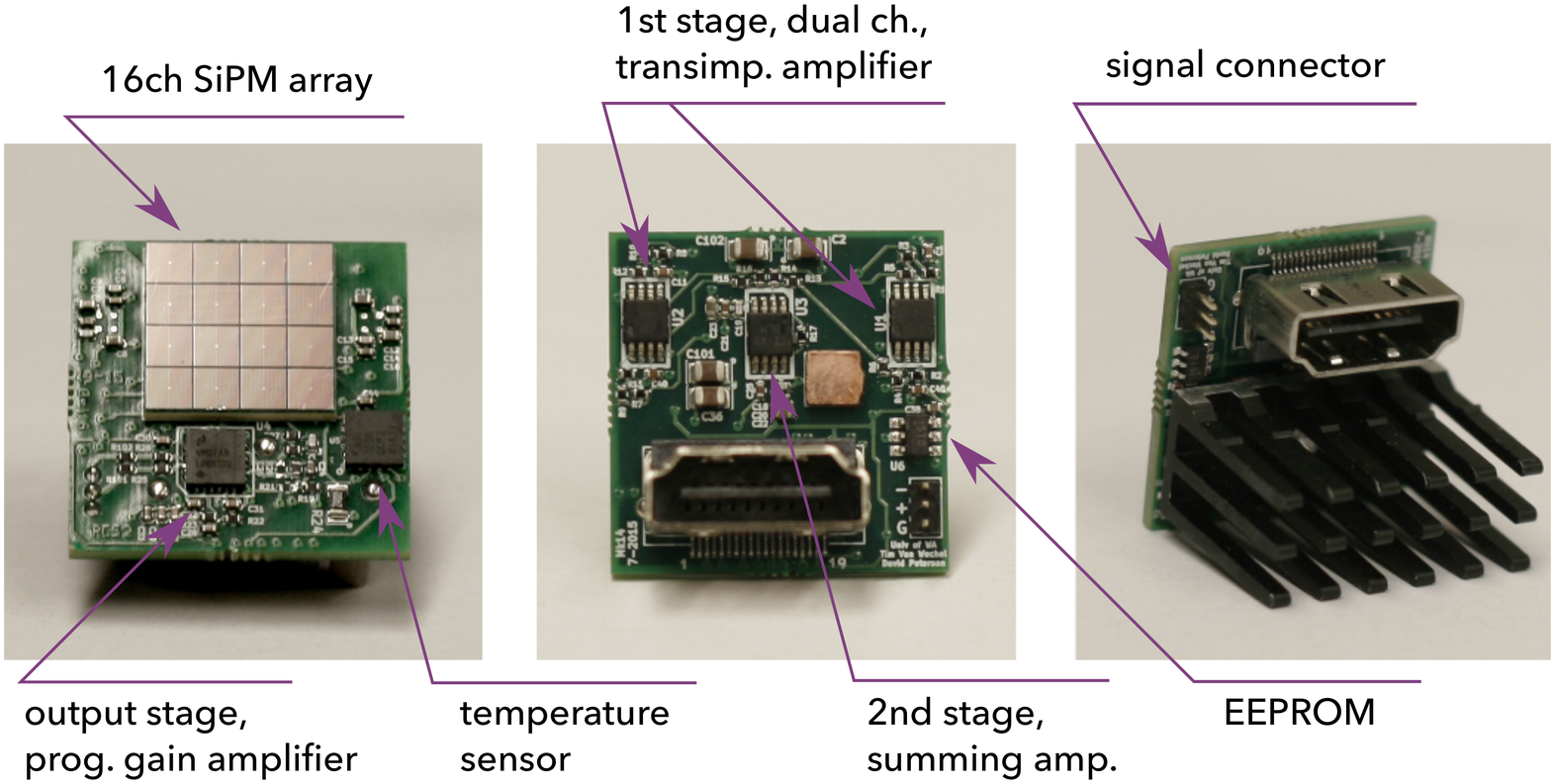}
\caption{The left and middle panels show the final amplifier board with the surface-mount 16-channel SiPM.
    The three-pin connector on the bottom right of the middle panel provides the AC-coupled differential voltage signal out. The common bias voltage in, the board low-voltage power, and the SPI lines used to regulate gain
    and readout temperature are supplied through the HDMI connector.
    The right panel shows the completed board after the heat sink has been attached. \label{fg:photo}}
\end{figure}

The practical realization of this circuit is shown in Fig.~\ref{fg:photo}, where both sides of the PCB are shown.  The outer board dimensions are constrained to be within the $25 \times 25$\,mm$^2$ calorimeter crystals onto which the SiPM ``front'' side of the board is attached using a high-index-of-refraction optical epoxy. The rear side of the board has a 3-pin output for the differential signal out, and an HDMI socket into which a standard HDMI cable is inserted.

\subsection{Operational point}
Our desired overvoltage operational point is determined primarily by optimization of the gain stability, as noise levels in our overvoltage range of interest are small, at the per mille level, compared to the expected signal. Of greater concern is large gain changes associated with charge depletion following large pulses in the SiPM and related rate-dependent gain shifts. By defining gain $G$ as the ratio of charge drawn through the SiPM to the number of incident photons\footnote{Note that this definition of gain folds in the changes in photon detection efficiency with overvoltage and therefore is nonlinear in the overvoltage.}, a pulse of a known light level will be proportional to the gain. The gain is a function of the bias overvoltage, $V$. Following a pulse of a fixed light level, a SiPM will see a temporary decrease in $V$ that is proportional to the size of the charge pulse and thus proportional to $G$. Therefore, to leading order, one would expect a fixed light level pulse to effect a gain perturbation of the following form:
\begin{equation}
 \frac{\delta G}{G_0} \propto - \frac{dG}{dV}r(t).
\end{equation}

In the above equation, $G_0$ is the unperturbed gain, and $r(t)$ is a recovery function varying between 1 at the time of the pulse and 0 at $t = \infty$.  The minus sign hints at the drop of gain following a pulse. This is a cumulative effect for multiple pulses and can therefore lead to rate-dependent gain shifts, which are a significant concern in the Muon \gm\ experiment. A key feature is that the relative gain drop is proportional to the derivative of gain with respect to overvoltage, a quantity that increases with increasing overvoltage for overvoltages reasonably close to zero.

Slow gain shifts can result from temperature changes. The SiPM breakdown voltage depends on temperature, so any uncompensated change in temperature will directly change the overvoltage and in turn the gain, which depends on overvoltage in a non-linear way. For infinitesimal changes and in the leading order of $dV$---in which the overvoltage is linear in temperature and the gain is linear in overvoltage, i.e., $G(V) = G(V_0) + \frac{dG}{dV} \delta V + \mathcal{O}(\delta V^2)$---one finds that small drifts in temperature $\delta T$ lead to changes in gain as follows:
\begin{equation}
	\frac{\delta G}{G_0} \propto - \frac{1}{G}\frac{dG}{dV} \delta T.
\end{equation}
In the case of temperature-dependent gain changes, the relative size of the effect is proportional to the logarithmic derivative of gain with respect to overvoltage, a quantity that decreases with increasing overvoltage. The same is true of drifts in bias voltage not related to pulses.

The recommended operational voltage is prescribed by the manufacturer using their method to obtain the desired gain. The method utilizes the first-order derivative of the current-voltage characteristics, see Fig 5 of the relevant patent application~\cite{sato2013photodiode}. The recommended voltage is prescribed as the breakdown voltage plus a fixed offset to achieve the desired multiplication factor. The breakdown voltage corresponds to the first inflection point on the current-voltage curve where most pixels turn into the Geiger mode. The fixed offset corresponds to the voltage difference between the first and second inflection points on the current-voltage curve. This choice of the offset balances the advantages of the high photodetection efficiency, and disadvantages of dark rate, crosstalk, and after-pulsing. The 16 channels of a SiPM array have different recommended operation points; however, for our purposes, the shapes of the current-voltage curves are similar enough for all of the channels of a SiPM array. Therefore, there is an effective current-voltage curve for a SiPM array on which the 16 voltage points recommended for the 16 channels lie. The 16 points are narrowly spread from each other in the linear region of the current-voltage curve. Owing to the uniformity of illumination, the 16 voltage points can be replaced by a single effective operational voltage, resulting in the effective gain of the SiPM array.

The manufacturer-recommended operational voltage is also our choice for the operational point in our application because it provides a sufficient gain, requires a temperature stability that can be met, and allows for the use of a single effective gain value for a whole SiPM array that scales with the bias voltage and temperature fluctuations in a predictable and near-linear manner.

\subsection{Production and Testing}
We contracted for 1400 SiPM amplifier boards to be fabricated and assembled commercially. It wasn't practically feasible to track the individual 16-channel SiPM arrays through the process of amplifier board assembly. The SiPM arrays were sorted into the groups of 10 SiPM arrays having similar average breakdown voltage $V_{\rm{bd}}$ specifications\footnote{Hamamatsu provides a measurement of $V_{\rm{bd}}$ for each of the 16 individual channels; the average value was used to sort the devices into groups of 10 for the purpose of amplifier board assembly.}.  The groups were shipped to the assembly house such that the completed boards would accurately reflect which SiPM group was attached to each board.

\begin{figure}[t]
\includegraphics[width=.45\textwidth]{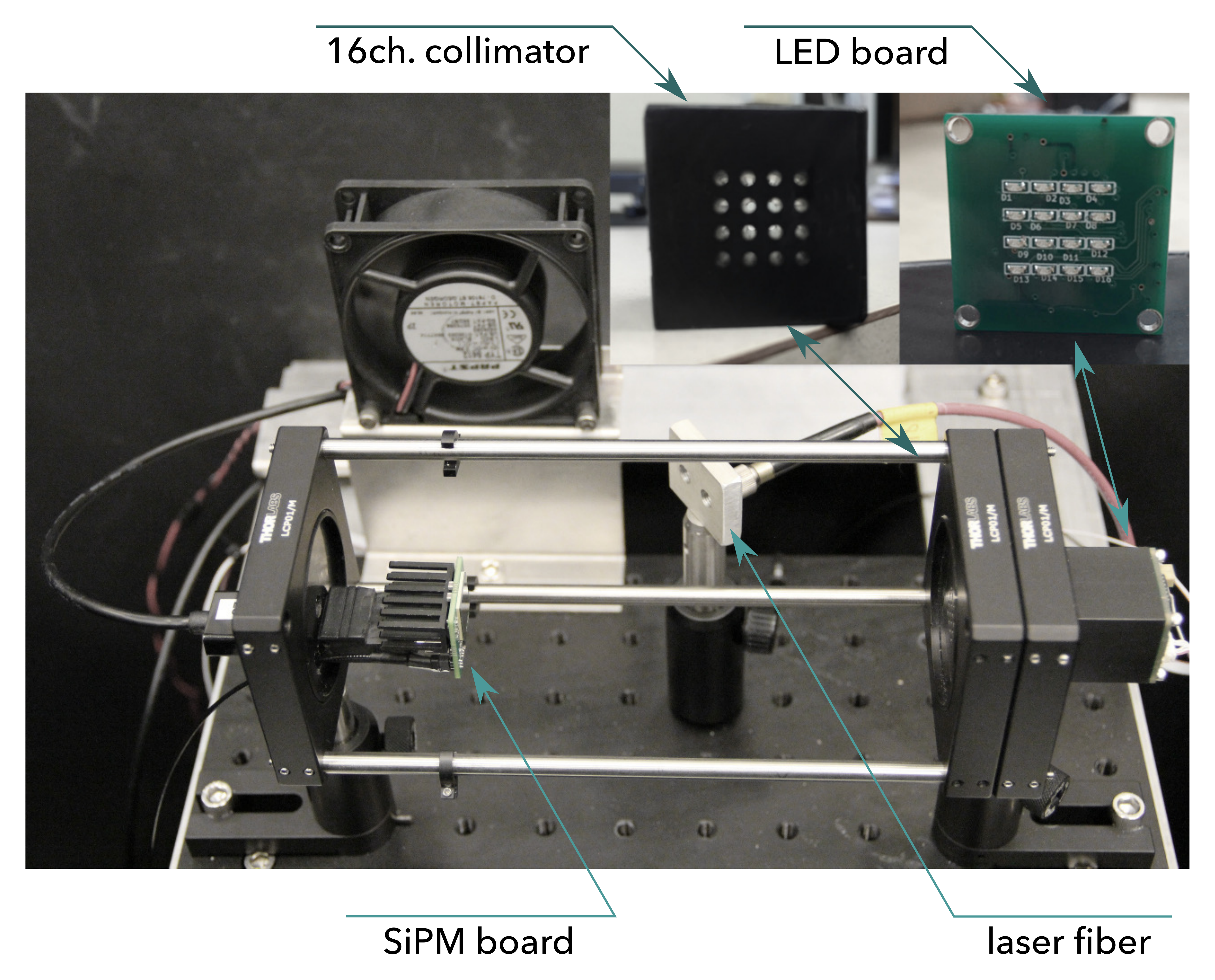}
\hfil
\includegraphics[width=.50\textwidth]{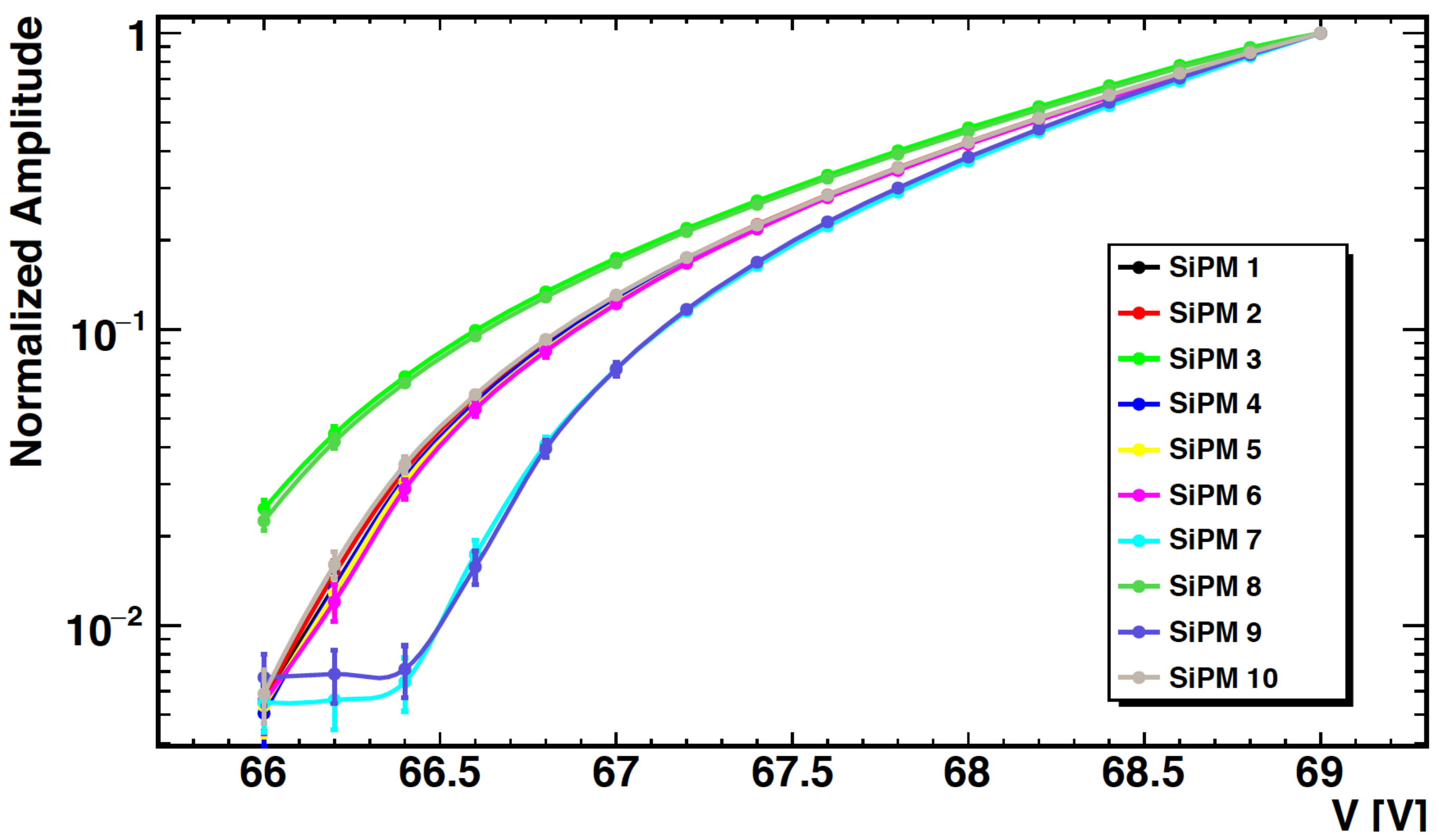}
\caption{Left: Elements of the QC station. The shown configuration features a fiber which illuminates the entire SiPM array from a laser light flash.  The left inset shows the 16-channel collimator that is used together with the 16 LEDs in the right inset to configure the station to flash the 16 channels individually. Right:
The normalized output amplitude for 10 SiPM arrays is plotted as a function of applied bias voltage. Three distinct groups of breakdown voltages exist in this group of 10 SiPM arrays.}
\label{fg:QCtests}
\end{figure}

Following assembly, a custom tailored heat sink was glued on the board, and then quality control (QC) tests were made.  A dedicated QC station was constructed to test each of the 16 channels, plus the PCB communications and the temperature readback.
As shown in the left inset of Fig.~\ref{fg:QCtests}, a 16-hole Delrin collimator can be aligned precisely with respect to the 16 channels of the SiPM to isolate each channel optically from the others.  An LED board having 16 SMD-type LEDs (right inset)  is mounted to the rear of the collimator and is programmed to flash each of the 16 LEDs one by one during QC tests. The summed SiPM signal represents the output of only one channel at a time. These data provide the verification that each of the 16 channels is functioning properly.

In addition, we also tested the temperature sensor, programmable gain amplifier (PGA), and the flash memory (EEPROM) installed on the board. Approximately 4\% of the boards failed the QC tests for various reasons, with most errors being recoverable by repairing or replacing various individual components, the external connector, or the seating of the heat sink, which accounted for 50\% of the problems. As an example of the power of the QC program, one board had five of the 16 channels not responding (for unknown reasons).  This defect would likely not have been noticed had we not flashed each channel individually.

In a second use of the QC station, the average breakdown voltage $V_{\rm{bd}}$ of the SiPM as a whole was measured with the aim of confirming the manufacturer's specification. The collimated LED system was replaced by a pulsed laser (red optical fiber in Fig.~\ref{fg:QCtests} left) that was positioned to illuminate the entire 16-channel SiPM at once. The bias voltage $V_{\rm{bias}}$ applied to the SiPM board was varied from 66\,V to 69\,V, and the output signal amplitude was plotted as a function of $V_{\rm{bias}}$.  Figure \ref{fg:QCtests} right shows the results for 10 SiPMs belonging to the same group; i.e., they were specified to have similar $V_{\rm{bd}}$ values. The curve for each SiPM is normalized to the amplitude at 69\,V. The SiPMs trend in three distinct groups as the bias voltage is lowered toward breakdown.

\subsection{SiPM bias load tester}
\label{sc:bias}
\begin{figure}[t]
    \includegraphics[width=0.95\textwidth]{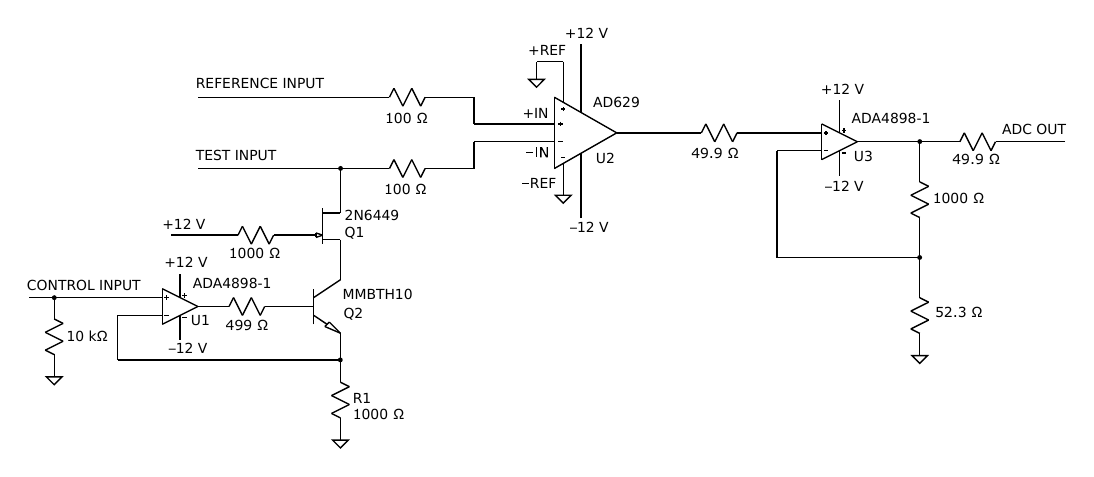}
\caption{Schematic of the SiPM bias load tester used to evaluate the performance of candidate bias supply units as a function of the rate and pattern of emulated events in the SiPM boards. \label{fg:loadtester}}
\end{figure}

The transient response of the bias voltage power supply is critical in our application.
The amount of light recorded by a SiPM decays exponentially with a half-life of $64\,\mu$s. For the purpose of testing and identifying a suitable power supply, a SiPM bias load tester (SBLT) was developed. It is a functional replacement of several SiPM boards from the power supply's point of view. The SBLT allows for a greater flexibility in loading the power supplies in patterns that might be difficult to achieve using the typical combination of an LED pulser and/or a short-pulse-duration laser. The simplified schematics of the SBLT is given in Fig.~\ref{fg:loadtester}.

The SBLT was developed to measure the transient recovery response of candidate bias supplies to simulated SiPM pulse current draws. The output voltage of a supply loaded by a voltage controlled current source is compared to that of a reference bias supply that is not loaded.  The difference between the two SiPM bias supply voltages is measured by amplifier U2, an AD629 high common-mode voltage difference amplifier.
The difference voltage at the output of amplifier U2 is referenced to ground and is amplified by op-amp U3 in a non-inverting configuration with a gain of 20.
The active components of the voltage controlled current source are op-amp U1 and transistors Q1 and Q2 connected in a cascode configuration, which improves the frequency response of the voltage controlled current source.
The SiPM load current flows through the cascode transistors.
The voltage drop across resistor R1 is proportional to the SiPM load current and is applied to the inverting input of amplifier U1.
The output of amplifier U1 drives the base of transistor Q2 of the cascode amplifier.  This feedback configuration forces the SiPM load current to be equal to the control input voltage divided by the value of resistor R1, assuming that the control input voltage is positive.
The control input voltage is generated using an Agilent arbitrary waveform generator programmed to simulate the current pulses that would be produced by the SiPMs in our application.

Current flowing through a resistor in series with a SiPM array results in a voltage drop across the resistor and effectively lowers the bias voltage experienced by the SiPM array. The decrease in bias voltage results in a lower gain of the SiPM. Some resistors are unavoidable for circuit stability reasons, and some are unavoidable from the power supply design perspective. While a charge buffer on a SiPM board can mitigate the reduction in gain for occasional load bursts, it is not a solution when the load rate is significantly higher than the transient response of a power supply.

We used the load tester, as well as subjecting an array of SiPMs to various light pulse sequences, to evaluate new commercial multi-channel bias supplies which had been developed for SiPMs.
All of the new products failed to provide stable voltage in a variety of tests. However, the BK Precision 9124A single-channel unit was successful. In all cases, the challenge was to re-supply charge fast enough to sustain a high-rate drain, such as when the SiPMs are pulses at a sustained rate of several MHz. The failure modes for the multi-channel units included that the output voltage went into oscillation, or that it sagged significantly under the load.  For those units, the advertised maximum current was no more than about 10\,mA. The BK supply, with a full-scale maximum voltage of 72\,V and a maximal output current of 1.2\,A, on the other hand, was successful in the high-rate and other tests.  The series resistance of the power supply was measured to be effectively $\approx 5\,\Omega$ for transient loads, and the measured transient response time was about $40\,\mu$s for transient loads at the few mA level.

The intermediate conclusion is that SiPMs are in principle capable of operation at very high rates without significant gain drop but only if extra care is given to selection of a suitable bias voltage power supply.  Further, it is important to achieve very low resistance of the complete power chain in series with the load. The desired quality of the power supply is its ability to maintain the voltage independent of the load.  Examples of test results are shown in Section~\ref{sc:rate}.

\section{Performance and Tests}
In this section, we report a number of results for the performance of the SiPM system described above. Characteristics of interest include the pulse shape, the time resolution, the calibration of fired pixels (or photo-electrons) to pulse height, the resolution of closely spaced pulses, the rate tolerance, and the gain stability as a function of rate. Measurements were carried out on the bench and using 3\,GeV electrons together with the full calorimeter crystal array.

During the development phase, a fast 405\,nm laser system was used to directly illuminate a SiPM over a range of rates and light level intensities. For dynamic gain stability tests, a background could be created by additionally firing a LED in known patterns with respect to the laser pulse. The LED driver was capable of creating a sharp pulse, whose SiPM response  closely resembled the laser pulse in both time and amplitude.

\begin{figure}[t]\center
    \includegraphics[width=0.75\textwidth]{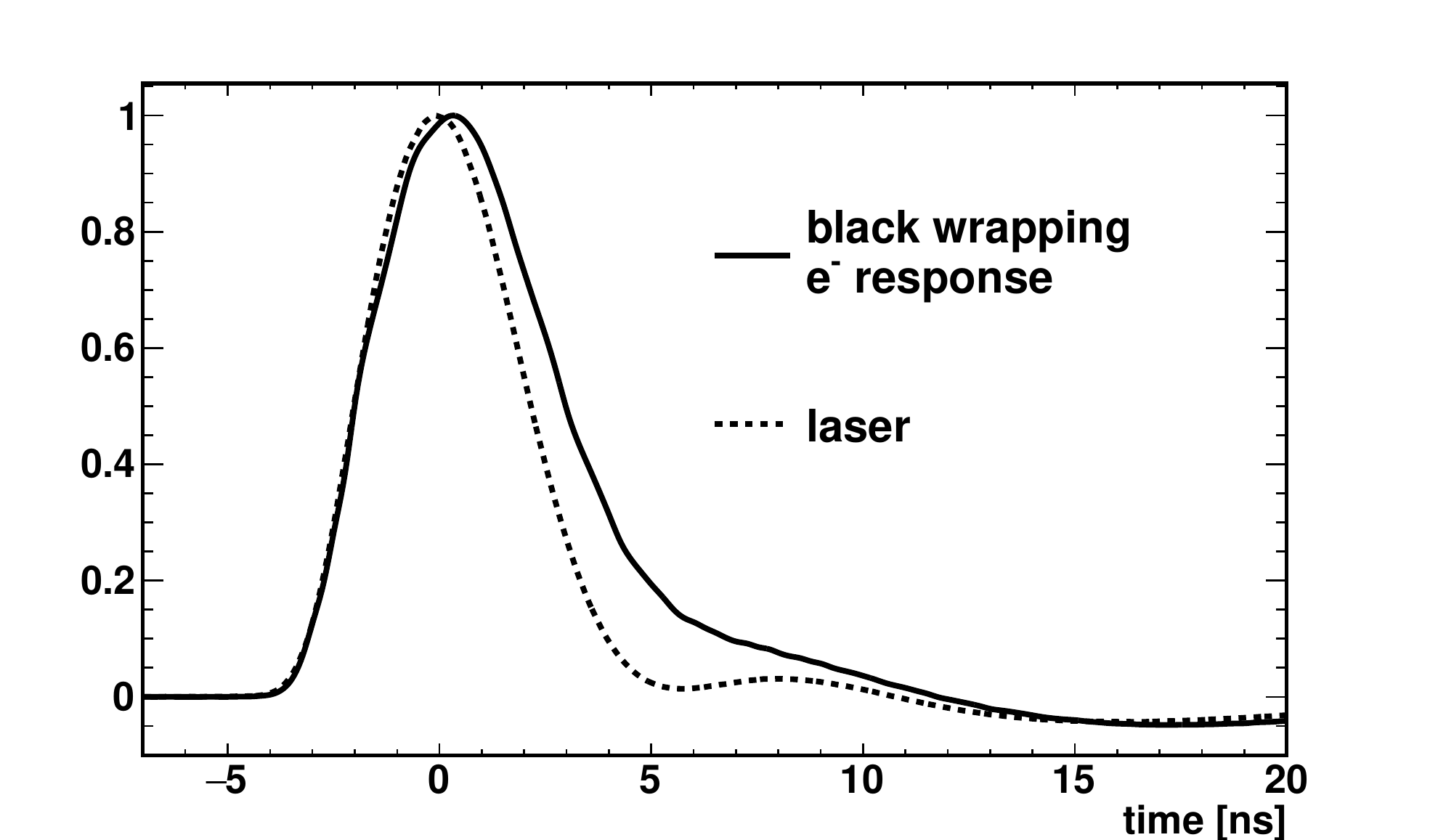}
    \caption{%
        Templates following a large number of laser (dashed) and beam (solid) events. The narrower shape reflects the near instantaneous light arrival from the  sub-ns laser photo-diode.  The wider shape is from 3\,GeV electrons striking normally in the center of a PbF$_2$ Cherenkov crystal and causing an electromagnetic shower.
        \label{fg:pulses}
    }
\end{figure}

The end-use configuration was tested using 3\,GeV electrons provided by the SLAC End Station Test Beam Facility. The results reported here are based on the final implementation of a complete 54-element  calorimeter station with individual SiPM readout of each \pb\ crystal in the array.  The signals from each of the channels were conveyed to custom 800\,MSPS, 12-bit digitizers developed~\cite{Chapelain:2015esj,Grange:2015fou} for the Muon \gm\ experiment.  The complete laser-based calibration system~\cite{Anastasi:2016efl,Anastasi:2016luh} was used to flash the front face of all crystals simultaneously at user-controlled rates and in specific pulse patterns.  The stability of the laser light was sufficient to allow sequences of calibration runs to be made in which the laser light intensity could be used to reliably extract the photo-electron to pulse height ratio for each SiPM and thus calibrate the individual gains independent of the beam.

\subsection{Pulse Shape and Calibration Procedure}
\label{sc:sc-calib}
A representative pulse shape response to the few-hundred ps wide laser pulse is shown in Fig.~\ref{fg:pulses}.  The same figure also shows the response to a 3\,GeV electron impinging centrally on the front face of one of the PbF$_2$ crystals. The pulse width from the beam is slightly wider, owing to the collection time of light from the electron shower.  The FWHM is $\sim 4.4$ for the laser shape and is $5.2$\,ns for the beam shape.  The smooth shapes are derived from a large number of individual pulses, which create  ``template'' patterns that are used in subsequent individual pulse fits.  A fit to a single digitized pulse requires only optimizing the time of the template because the amplitude is fixed from the integral of the pulse area. Importantly, the pulse shape is not a function of the number of pixels fired or of the rate.  However, it is clearly a function of the light source, which in the application will be distinguished between beam or laser owing to external trigger tags.

\begin{figure}[t] \center
    \includegraphics[width=0.75\textwidth]{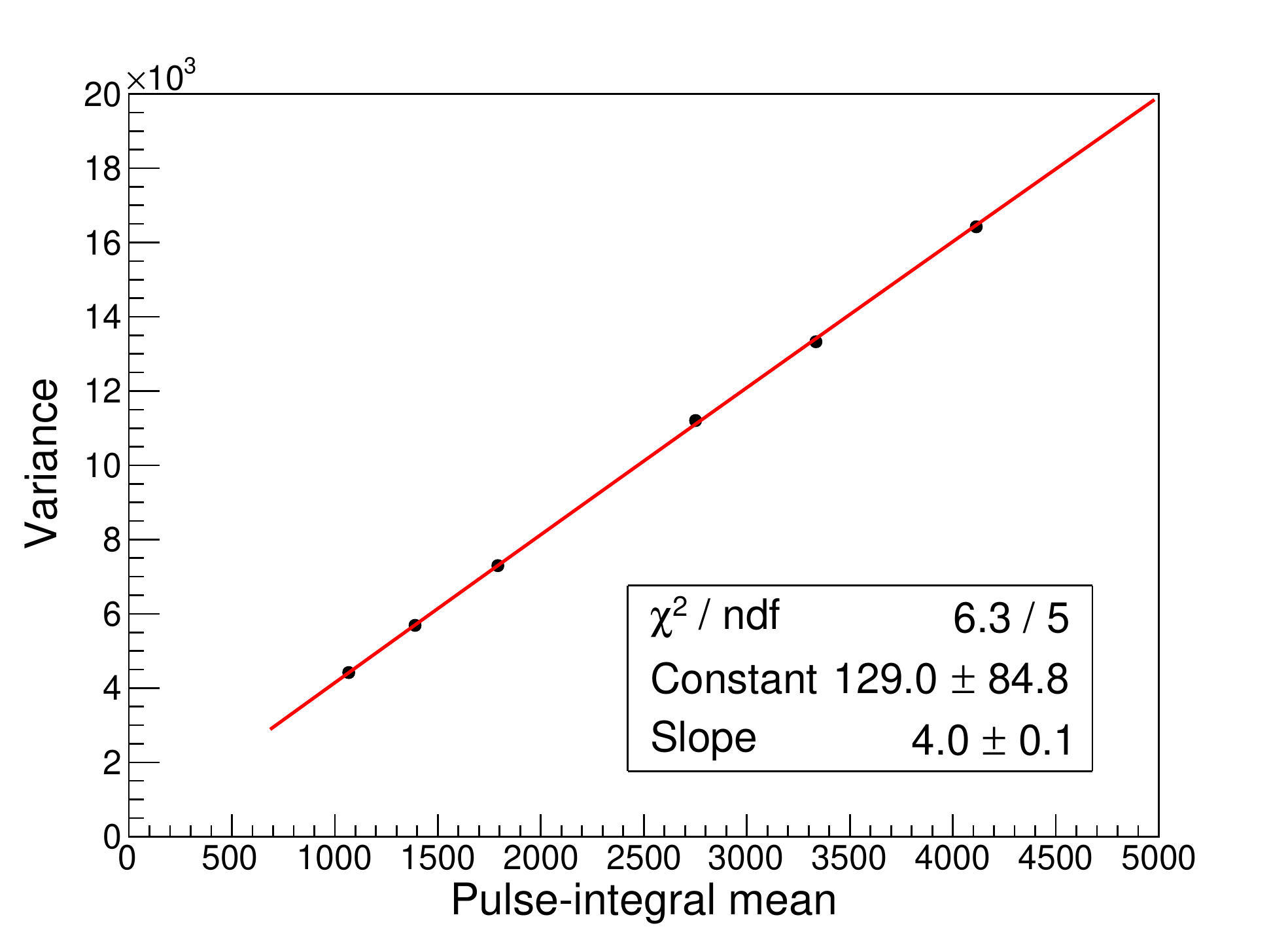}
    \caption{%
    Plot of variance $\sigma^2$ versus pulse-integral mean $M$ of a distribution of fitted laser pulses on a SiPM.  The discrete mean values are based on using a multi-step filter wheel to attenuate the light.  The sequence of runs begins and ends in the open position of the filter wheel, corresponding to the upper right most point. The inverse of the fitted slope corresponds to the PE/M and the good linearity implies that the variance entirely depends here on the statistics of the number of pixels fired on each event.\label{fg:calibration}
    }
\end{figure}

Slight differences in pulse shape can be observed and are expected, owing to the arrival time of photons at the SiPM surface from showers initiated from electrons over a large angular range. In those cases, Monte Carlo anticipates the photon arrival profile, which is reflected in a slight increase in the pulse width. In our implementation, the crystals are wrapped in black Tedlar$^{\circledR}$ foil to minimize the fraction of reflected --- and therefore late-arriving --- photons at the SiPM surface.

The calibration of photo-electrons per mean signal pulse integral ($M$) is made by the method described in Ref.~\cite{Fienberg:2014kka}.  For a Poissonian distribution of photons from the light source,  the number of PE is obtained from the ratio $M^2/\sigma^2$, where $M$ is the mean of a distribution of pulses, and $\sigma$ is the width of that distribution. The calibration procedure uses an automatic sequence of 10,000 pulse runs.  Each run uses a unique setting of a neutral density filter wheel to control the light intensity in finite steps. Figure~\ref{fg:calibration} shows a typical example of this process.

\subsection{Timing resolution}
\begin{figure}[t] \center
    \includegraphics[width=0.75\textwidth]{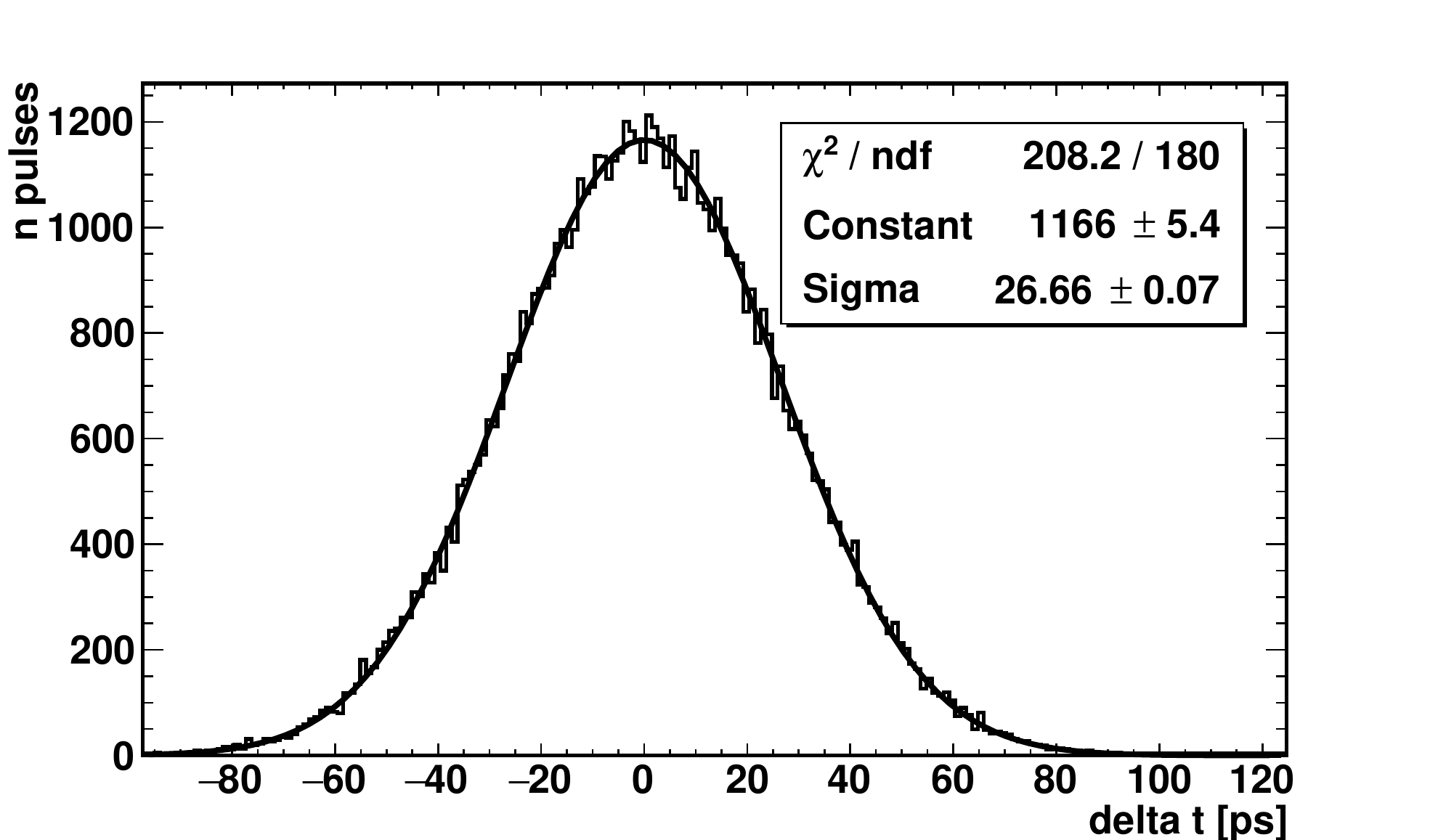}
    \caption{Time difference in ps for two SiPMs each illuminated from the same laser pulse with approximately 2000 PE/pulse. The 19\,ps time resolution of individual channels is a factor  $\sqrt{2}$ smaller than the quoted width. \label{fg:timing}
    }
\end{figure}

Given the stable pulse shape and accurate template fitting procedure, the timing resolution can be deduced from either the laser or beam events.  The controlled response with the laser is more straight-forward because nearly equal amplitude light pulses are injected simultaneously into all crystals.  The difference between the fitted time of two SiPMs recorded within the same digitizer module is shown in Fig.~\ref{fg:timing}. The width implies that the individual channels have a timing resolution of 26.7\,ps / $\sqrt{2}$ = 19\,ps.  In this measurement, the light intensity in each crystal created approximately 2000 fired pixels.

\subsection{High- and Variable-Rate Performance Tests}
\label{sc:rate}

A critical test of either a SiPM or a conventional PMT in a high-rate experiment is the stability of the response to a high sustained average rate of light pulses and the short-timescale gain recovery following a significant pulse.  It is well known that special care must be taken to maintain high voltage for PMTs in high-rate applications because the current drain in the capacitors of the base typically leads to pulse sagging or other pulse-shape distortions.  How the gain might be affected by available charge has been discussed briefly in Section~\ref{sc:bias}.  The issue is related to the average number of pixels fired in an event as well as to the time scale to re-supply the charge that has been drained by the fired pixels on the leading pulse. There are several relevant time scales and pulse patterns that we tested.

\begin{figure}[t] \center
    \includegraphics[width=0.62\textwidth]{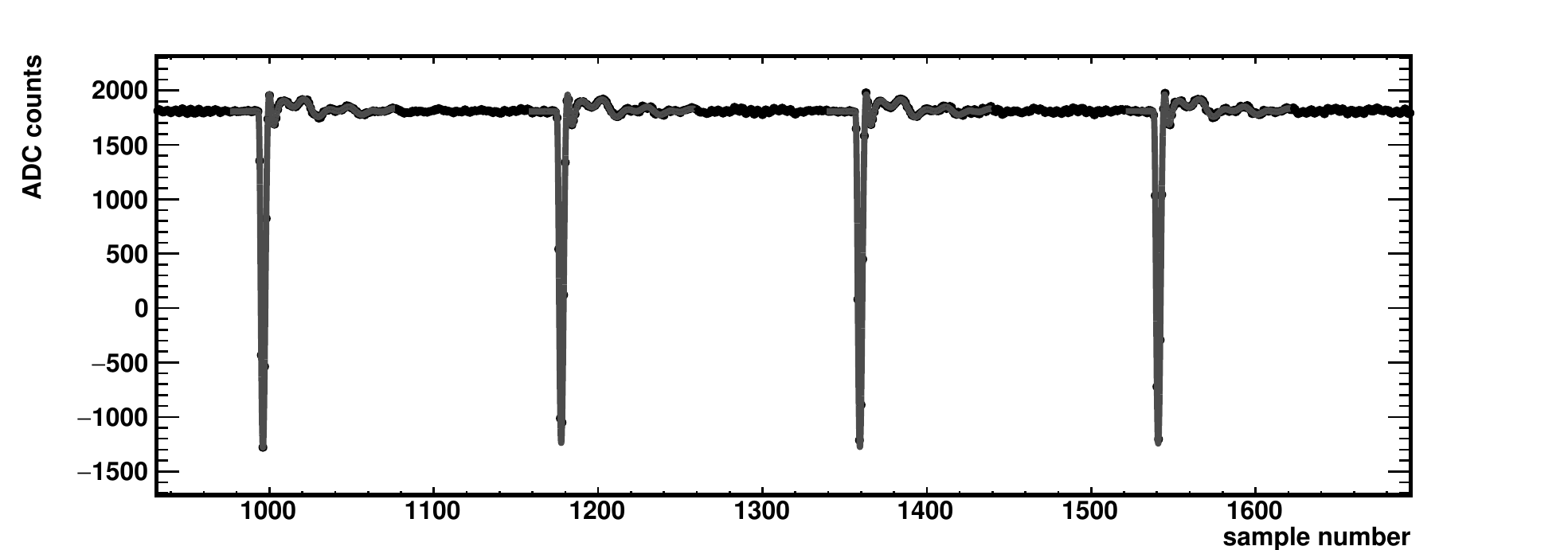}
    \hfil
    \includegraphics[width=0.32\textwidth]{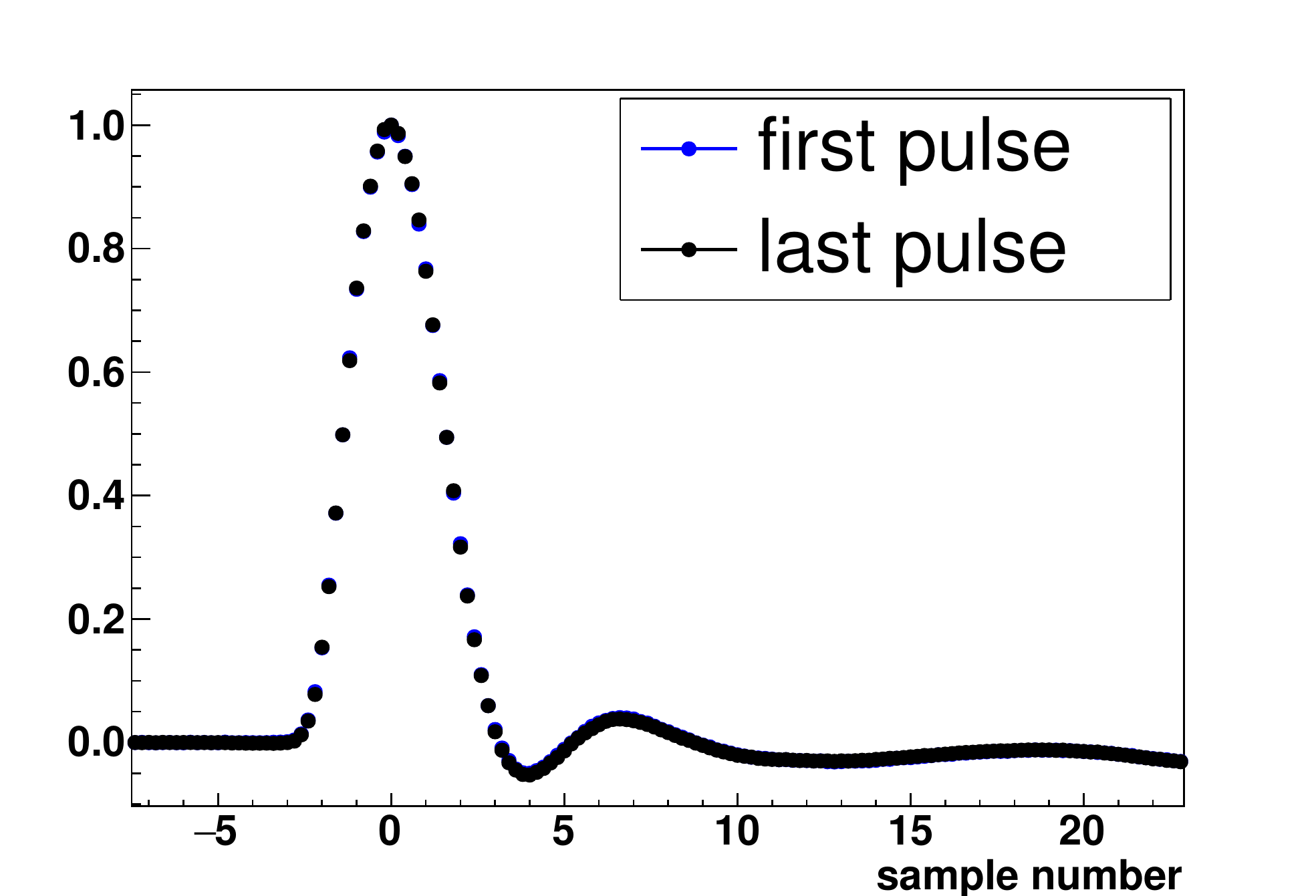}
    \caption{%
        Response of the SiPM array to a direct uniform illumination of its surface using the Picoquant laser operating at the rate of 5\,MHz. Left:
        a digitized trace is fitted using a template created from the first pulse of the train.
        Right:
        the first and last pulses of the pulse sequence are overlaid on top of each other. Most points are indistinguishable.
        \label{fg:high-rate}
    }
\end{figure}

Consider first the response to a high sustained pulse rate. We used the laser system described above to fire a train of pulses at the rate of 5\,MHz into a single SiPM. Figure~\ref{fg:high-rate} left shows a trace recording of this sustained sequence of pulses, digitized at 800 MSpS. The first pulse of the pulse sequence was used to create a template for the fitter. All pulses fitted successfully using this template. The pulse shape is unchanged versus rate, showing no intrinsic signs of distortion, which is shown in Figure~\ref{fg:high-rate} right, where the first and last pulse of a pulse sequence are overlaid on top of each other. Critical to this test is a bias supply capable of delivering a high-enough current to maintain the voltage independent of the load.



\begin{figure}[t] \center
    \includegraphics[width=0.45\textwidth]{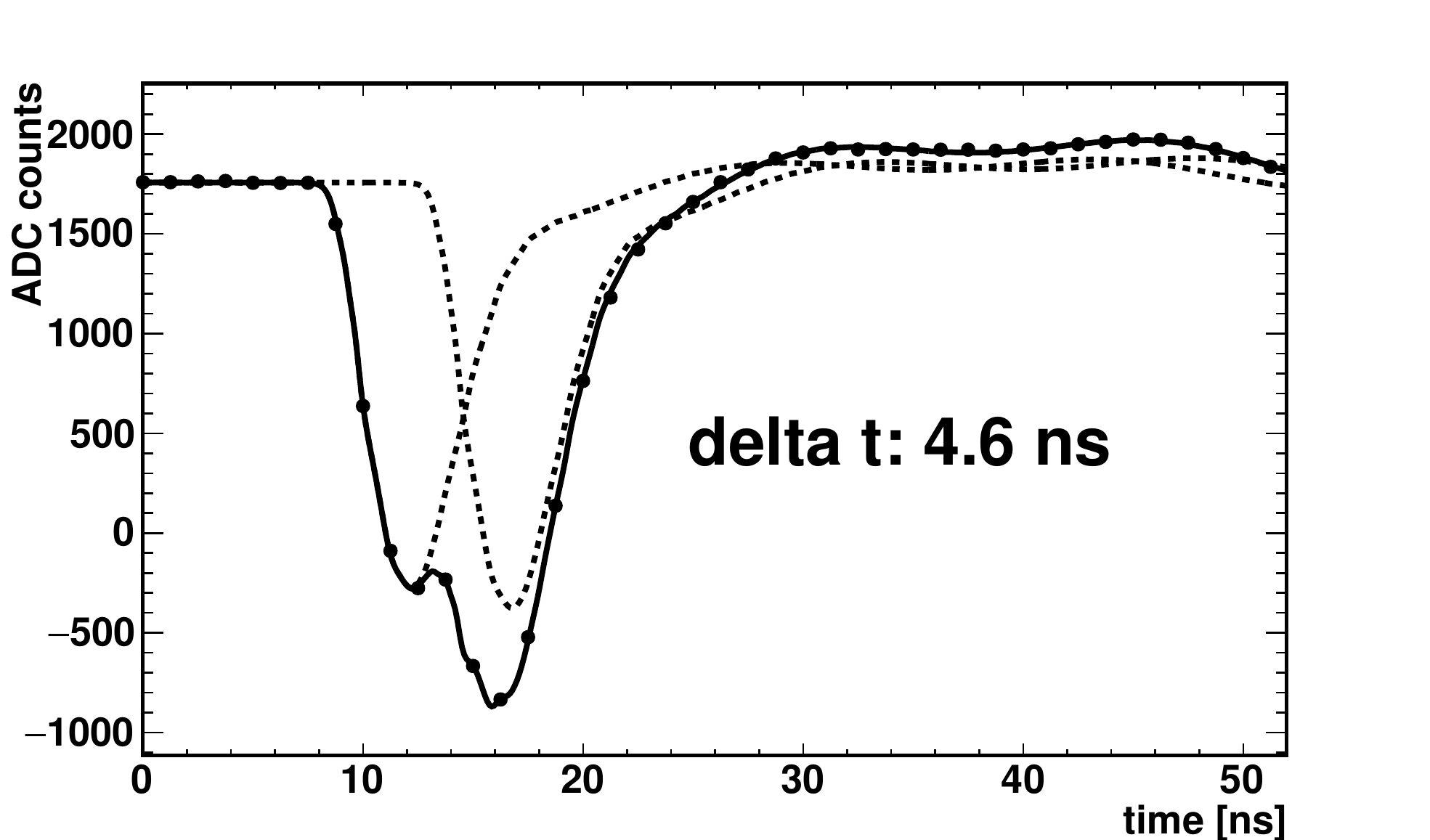}
    \hfil
    \includegraphics[width=0.45\textwidth]{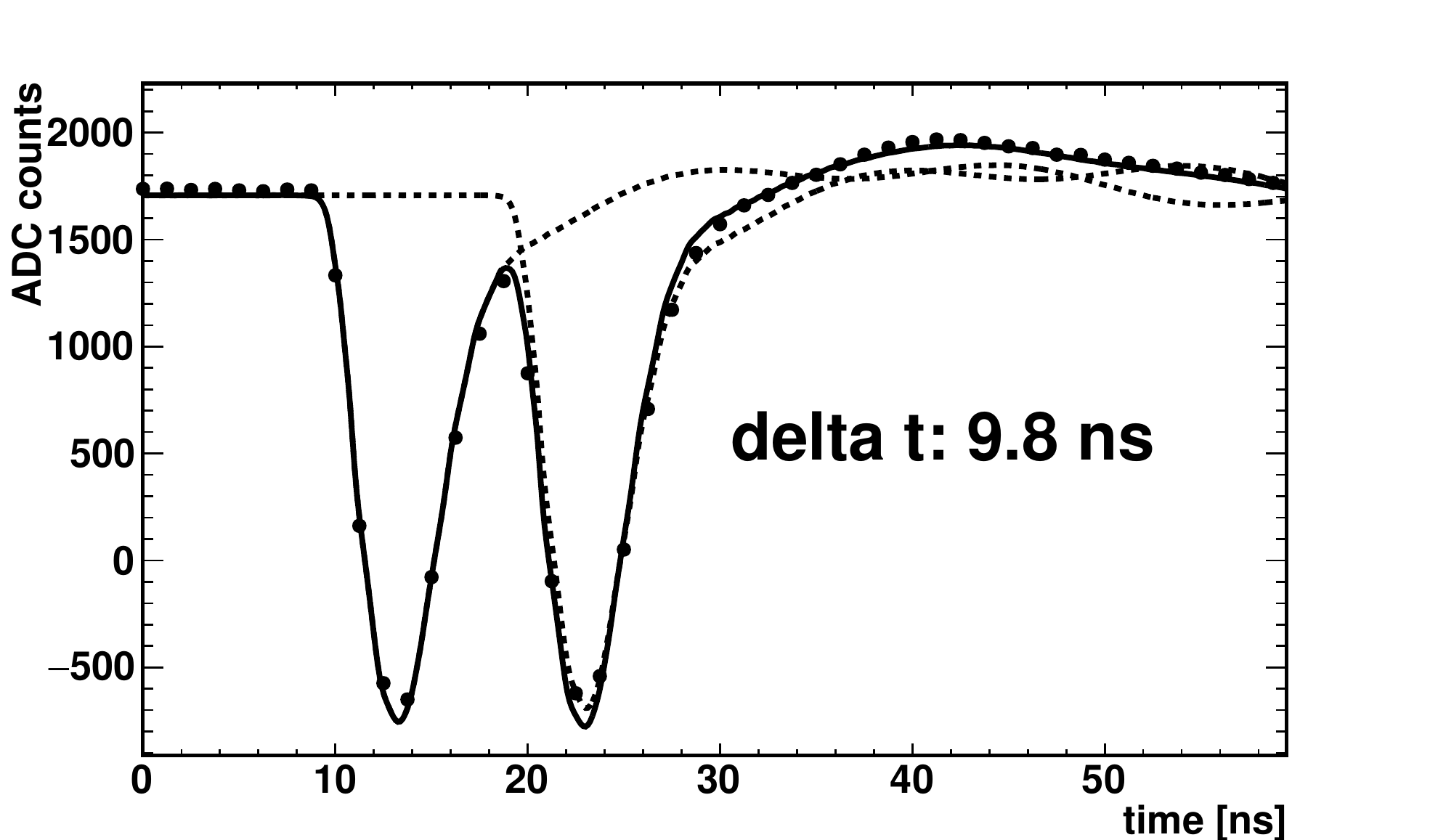}
    \caption{Two 3-GeV electrons striking near the center of the same crystal with time separations of 4.55\,ns (left) and 9.8\,ns (right). Depending on the exact impact position, each pulse corresponds to approximately 2000\,PE.  The two pulses are easily resolved in both cases using the template fit. A small and predictable correction for the trailing pulse gain can be made based on the study of the short-term gain recovery.\label{fg:2-pulse}
    }
\end{figure}

The next example is related to the gain immediately following a significant pulse; i.e., on the time scale of several tens of ns, which is relevant for resolving two-pulse pileup events such as those shown in Fig.~\ref{fg:2-pulse} where one needs to obtain correct times and energies of the pulses. The same time scale is typical for the SiPM chip to charge up, i.e., for the charge to transport from the local power supply on the SiPM board, formed by large capacitors, to the SiPM chip itself. When a SiPM pixel fires, it delivers charge, which is replenished afterwards.
The delivered charge comes from all available sources, particularly the SiPM array itself. The SiPM array behaves as a charged capacitor; i.e., the bias voltage experienced by the SiPM array drops momentarily before the charge is recovered. Figure~\ref{fg:spice-pulses}, which plots the SPICE model pulse shapes throughout the circuit evolution, also shows the predicted voltage drop on the buffer capacitor as a function of time following a pulse.  The scale is in arbitrary units; this voltage drop is tiny.

As an example, suppose that 1600 pixels fired simultaneously on the SiPM board, 100 pixels in each of the 16 channels. The charge delivered by a fired pixel is $1.5 \times 10^6$ electrons, for a total charge for each channel of 24\,pC.  This results in a voltage drop of 96\,mV, taking into account the 250\,pF capacitance of each of the 16 channels.  The relative voltage drop is 4\% compared to the typical operating overvoltage bias of 2.5\,V.

The charge is replenished from the capacitors on the SiPM board. The only available path for the charge to the SiPM pixel goes through its quenching resistors. The product of the pixel capacitance, 55.8\,fF, and the value of the quenching resistor, 100\,k$\Omega$, sets the expected timing constant for the charge recovery of about $\sim6$\,ns. To verify the charge recovery model and better understand the details of the charge recovery mechanism, the following exercise was performed.

\begin{figure}[t] \center
    \includegraphics[width=0.75\textwidth]{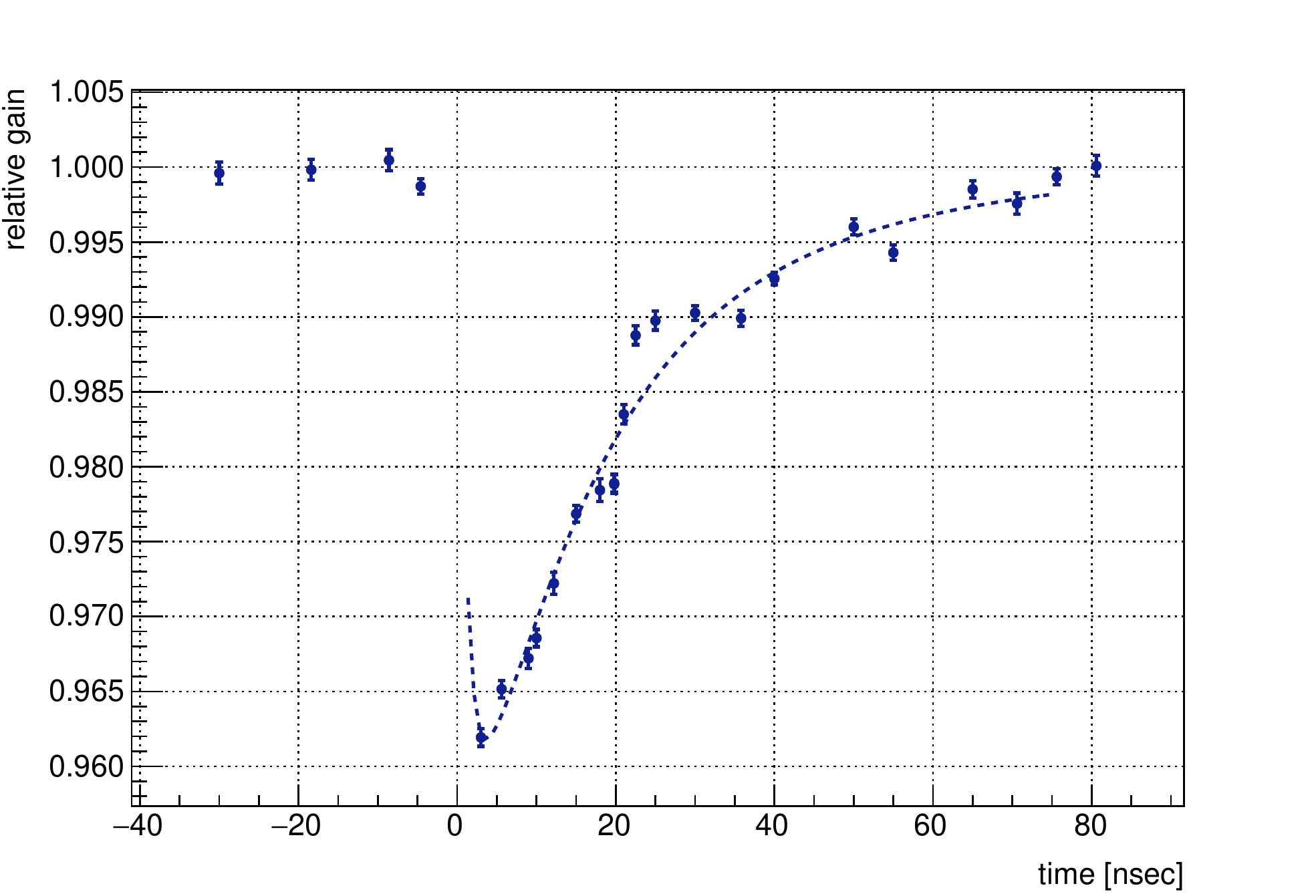}
\caption{Relative gain of the laser pulse with respect to the reference LED pulse versus a fixed time interval between the two pulses. \label{fg:shortterm}}
\end{figure}

In the study here, the SiPM board was fired using a LED pulse, and the process of charge recovery was probed using the laser. The LED pulse shape was tuned to match the laser pulse shape as closely as possible. The amplitude of both pulses corresponded to about fired 1600 pixels per SiPM board. The laser pulse was scanned from $-40$\,ns to $+80$\,ns with respect to the LED pulse. Each measurement consisted of two steps: 1) in the first step, only the laser is fired, 2) in the second step, both the LED and laser are fired.
The pulse integrals were extracted by a custom two-source template fit, and the ratio of the laser pulse-integrals was calculated as the second step compared to the first step, i.e., the SiPM undergoing recovery compared to the unloaded SiPM. Assuming that the number of photons fired by the laser was the same in both steps --- a good assumption --- the ratio directly reveals the gain drop as a function of pulse-time separation. The measurement was repeated many thousand times, and the ratios were averaged to obtain a single point in Fig.~\ref{fg:shortterm}. This measurement method is robust with respect to the fluctuations in laser intensity and temperature.

The results of this gain study indicate two components of the gain recovery time response.  A representative fit is shown in Fig.~\ref{fg:shortterm} with 9- and 29-ns recovery time constants and relative amplitudes of 1.7 to 1, respectively.  Note that this fit is merely a guide to the eye.  The faster component is to be compared with the expected value of $\sim6$\,ns. The difference can be explained by the extra parasitic impedances of the PCB and SiPM chip, which are unaccounted for in our simple model. The slower component is typically attributed to the bulk properties of the SiPM chip, which is a satisfactory explanation for our purposes. In our application, the measured gain recovery allows us to correctly separate pileup events in the energy domain. The predictive drop measured here can also be used to correct the gains of trailing pulses in the reconstruction phase.


The next test is somewhat application specific, but it may be of general interest as an example that required tweaking some component values on the PCB to minimize gain variance.
In the Muon \gm\ experiment, the storage ring is filled with muons, and observation of their subsequent decays occurs for  $700\,\mu$s, with a time-dilated muon lifetime of $64\,\mu$s.  Immediately after injection, the instantaneous rate on many of the detectors will exceed several MHz, but then it drops by more than four orders during the measuring period.  From the SiPM perspective, an average physics event in a single crystal will fire 250 pixels of a SiPM, and there will be approximately 72 events such events during a measuring period, distributed exponentially in time. This kind of transient load will result in a gain sag, which must be minimized and also precisely determined in our application.

The local power supply on a SiPM board is formed by the capacitance of $5~\mu$F.  The transient load of a SiPM array amounts to 5\,nC for the typical SiPM gain of $1.5 \times 10^6$ electrons per pixel fired. Without any external bias voltage power supply, the delivered charge lowers the bias voltage by 1\,mV at the end of the $700~\mu$s period. For the typical bias over-voltage of 2.5\,V, the relative voltage drop would be $4 \times 10^{-4}$,  well within the specification on the acceptable gain drop. A detailed SPICE simulation, including a bias voltage power supply and power distribution lines, predicts the bias voltage drop on a SiPM array of $300~\mu$V. For an external bias voltage power supply to help with the transient load, its transient response needs to be faster than the load. The stability of a SiPM board also requires a $2\,\Omega$ resistor in series with the power supply to decouple the cable. The measured transient response of the BK power supply was on the order of $40~\mu$s  in our setup.

To test the gain stability in this situation, the laser was programmed to fire in a pseudo-random pattern with an instantaneous rate that followed an exponential decay with time constant $\gamma \tau_\mu$. The spectrum of pulses was analyzed for pulse height
stability under the assumption of a stable laser light level throughout the pattern. The validity of the assumption was verified by a neutral density filter which varied the load on the SiPM but preserved the load on laser. The observed gain drop scaled exactly as expected for a given filter value.

\begin{figure}[t] \center
    \includegraphics[width=0.75\textwidth]{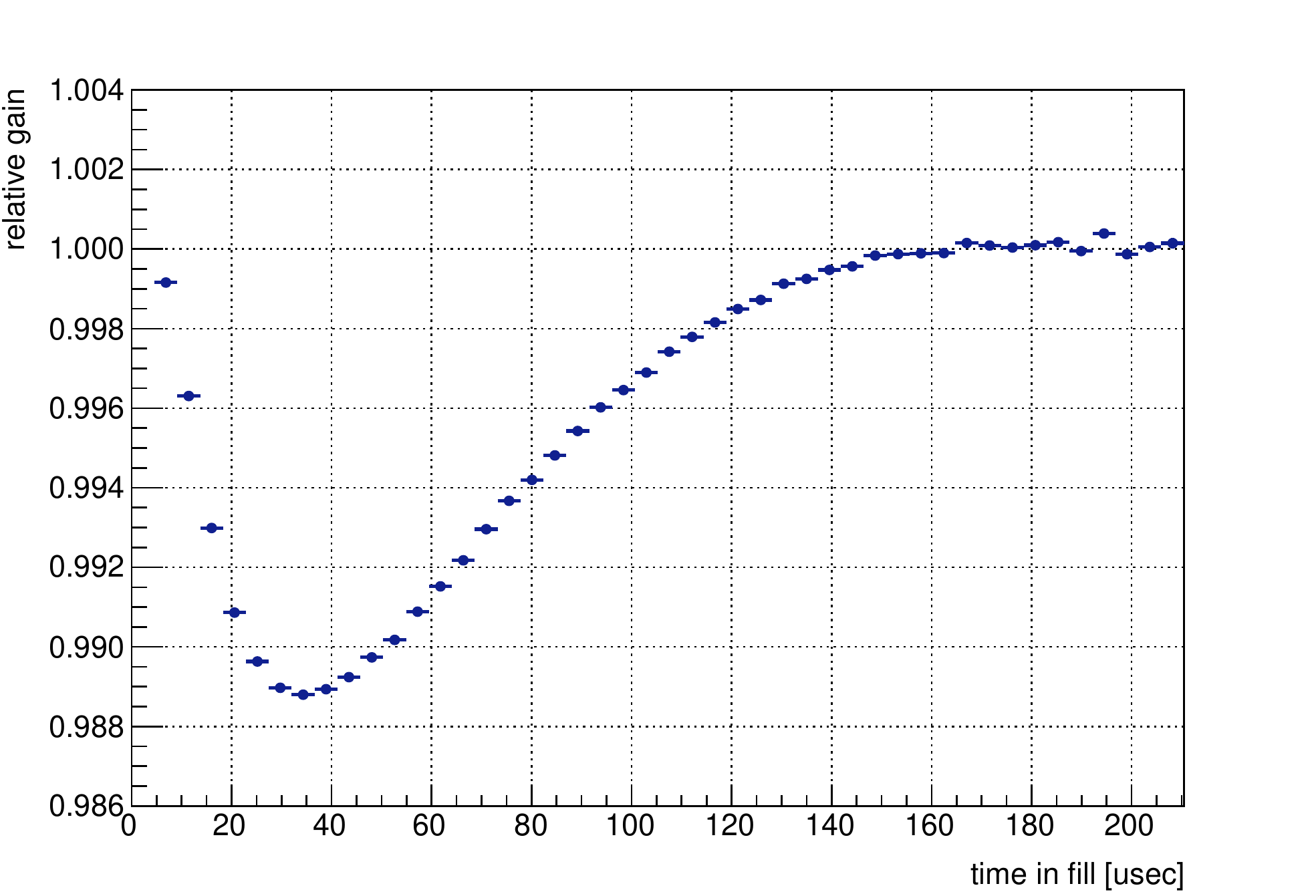}
    \caption{%
        Average gain as a function of time with an initial rate of 2.1\,MHz, 1500 PE pulses, and a rate proportional to $exp(-t/\tau)$ with time constant $\tau = 64\,\mu$s.  The integrated current drawn is higher than we expect in the Muon \gm\ experiment for the hottest crystal by a factor of 10. The slight droop and subsequent recovery of the gain is affected by a combination of the bias voltage supply and the buffer capacitance.
        \label{fg:96pulses}
    }
\end{figure}

Figure \ref{fg:96pulses} shows the results of a test with an instantaneous load approximately 10 times greater than expected in the experiment. The observed largest gain sag of 1.1\% will be reduced proportionally in the application.  Knowledge of the full gain curve will be made to sub-per-mille precision using a custom laser calibration system.

\section{Conclusions}

We have described our development of a custom amplifier board that supports a large-area, 16-channel SiPM from Hamamatsu.  While the design was driven by the end-use calorimetry requirement for the new \gm\ experiment at Fermilab, the desired performance characteristics are relatively common in many nuclear and particle physics applications.  They include:  operation in high magnetic fields, good (near) linear response for pulses of hundreds to thousands of photo-electrons, operation at MHz rates, short-duration ``PMT-like'' pulse shapes, and a high degree of gain stability. The final design PCB, together with careful attention to the external bias supply, achieves these goals.
Notably, we measure $\sim5\,$ns pulse widths and achieve $\sim20$\,ps time resolution in the tests described.

For ease-of-use, the PCB also includes a variable gain amplifier, an on-board temperature readout, and a built-in EEPROM for identification of the unit and storage of gain settings.  The low voltage, bias voltage, and communication lines are all conveyed using commercial HDMI cables.  The differential signal output is conveyed by a custom Samtec cable.  The PCB components and the cabling are non-magnetic.\footnote{Owing to the high sensitivity of any materials placed in or near the Muon \gm\ storage ring that could perturb the uniformity of the magnetic field, we had to remove by hand the HDMI ferromagnetic cable end clips.  Additionally, we had to select non-magnetic capacitors for the PCB.}  At this time, we have completed the production and testing of 1400 SiPM / amplifier boards.  Lengthy running of subsets of them over weeks of operation has shown no operational failures.

\acknowledgments

We thank A. Para for excellent advice on evaluating and choosing SiPM products and C. Hast for hosting us during the SLAC test beam measurements.  We acknowledge earlier contributions to our SiPM testing and development from L.\,P.\,Alonzi, B.\,Kiburg, and P.\,Winter.  This research was supported by the U.S. National Science Foundation MRI program (PHY-1337542), by the U.S. Department of Energy Office of Science, Office of Nuclear Physics under Award Number DE-FG02-97ER41020, by the Istituto Nazionale di Fisica Nucleare (Italy), and by the EU Horizon 2020 Research and Innovation Programme under the Marie Sklodowska-Curie Grant Agreement No. 690835.

\bibliography{jinst_sipm}

\end{document}